\documentclass[runningheads,a4paper]{llncs}
\usepackage[T1]{fontenc}
\usepackage{graphicx}
\usepackage{booktabs}
\usepackage{multirow}
\usepackage{xcolor}
\definecolor{indigo}{HTML}{4338CA}
\usepackage{enumitem}
\usepackage{amsmath}
\usepackage{amssymb}
\usepackage[caption=false]{subfig}
\usepackage{tikz}
\usetikzlibrary{positioning, arrows.meta, shapes.geometric, fit, calc, backgrounds}
\usepackage{cite}
\usepackage{hyperref}

\expandafter\def\expandafter\UrlBreaks\expandafter{\UrlBreaks\do\/\do\-\do\_\do\a\do\b\do\c\do\d\do\e\do\f\do\g\do\h\do\i\do\j\do\k\do\l\do\m\do\n\do\o\do\p\do\q\do\r\do\s\do\t\do\u\do\v\do\w\do\x\do\y\do\z\do\A\do\B\do\C\do\D\do\E\do\F\do\G\do\H\do\I\do\J\do\K\do\L\do\M\do\N\do\O\do\P\do\Q\do\R\do\S\do\T\do\U\do\V\do\W\do\X\do\Y\do\Z\do\0\do\1\do\2\do\3\do\4\do\5\do\6\do\7\do\8\do\9}

\begin{document}

\title{FORGE: Multi-Agent Graduated Exploitation\\and Detection Engineering}
\titlerunning{FORGE: Multi-Agent Graduated Exploitation and Detection Engineering}

\author{Farooq Shaikh}
\authorrunning{F. Shaikh}
\institute{Dynatrace, Vienna, Austria\\
\email{farooq.shaikh@dynatrace.com}}

\maketitle

\begin{abstract}
Vulnerability disclosure volumes now far exceed organizational assessment capacity, yet three adjacent research communities (proof-of-concept generation, vulnerability prioritization, and detection rule engineering) operate largely in isolation. Existing automated exploit generation systems report binary pass/fail outcomes, discarding partial progress and producing no signal for the other two communities.

This paper presents FORGE, a multi-agent system that bridges these three silos through \emph{graduated exploitation depth}. Five specialized agents (Intel, Generator, Planner, Exploit, and Detector) execute in a fixed pipeline that (1)~generates targeted vulnerable applications from CVE metadata, (2)~conducts coached, multi-turn exploitation assessed by an LLM-primary oracle on a four-level taxonomy (L0: no evidence through L3: full compromise), and (3)~produces Sigma and Snort detection rules grounded in OpenTelemetry exploitation traces. Graduated depth is the bridging mechanism: deeper exploitation yields richer behavioral traces for detection engineering, while depth data across scoring bands provides ground truth for prioritization validation. A tiered knowledge architecture accumulates intelligence across assessments, transferring build and exploitation experience to subsequent CVEs.

Evaluation on 603 CVEs from the CVE-GENIE dataset achieves 67.8\% end-to-end L1+ exploitation at \$1.50 per CVE across eight languages and 187 CWE types. Exploitation rates remain near 68\% regardless of EPSS or CVSS band, indicating that pattern-level reachability is orthogonal to metadata-based prioritization. Detection rules from L2+ exploitation achieve significantly higher span-normalized grounding than L1-derived rules ($p{=}0.035$), and 93.4\% of generated Snort rules produce zero false positives against a synthetic benign corpus.

\end{abstract}

\section{Introduction}
\label{sec:introduction}

Vulnerability disclosure rates now outpace human analysis by a widening margin: the Common Vulnerabilities and Exposures (CVE) program published over 48,000 entries in 2025~\cite{cve2025metrics}, yet security teams continue to rely on metadata-driven triage that has never been validated against actual exploitation attempts~\cite{jacobs2023epssv3,parla2024epss}. Common Vulnerability Scoring System (CVSS) severity scores and Exploit Prediction Scoring System (EPSS) exploitation probabilities, the two most widely deployed prioritization instruments, both operate on surface-level indicators rather than ground-truth exploitability data. EPSS uses exploit-code existence as a binary feature but never verifies whether that code achieves exploitation~\cite{jacobs2020epss,jacobs2023epssv3}; EPSS scores are frequently unavailable or low at the time vulnerabilities enter the Cybersecurity and Infrastructure Security Agency (CISA) Known Exploited Vulnerabilities catalog~\cite{parla2024epss}; and CVSS, Stakeholder-Specific Vulnerability Categorization (SSVC), and EPSS produce conflicting assessments on identical CVEs~\cite{koscinski2025conflicting}.

Automated exploit generation offers a promising path toward ground-truth exploitability data at scale. Fang et al.~\cite{fang2024exploit} demonstrate that GPT-4 agents can exploit 87\% of 15 one-day CVEs given their descriptions, and subsequent multi-agent systems (including CVE-GENIE~\cite{ullah2025cvegenie}, CVE-Bench~\cite{zhu2025cvebenchagents}, and PoCGen~\cite{simsek2025pocgen}) scale to hundreds of CVEs using hierarchical agent architectures~\cite{zhu2024hptsa}. However, all such systems report binary outcomes: a CVE either has a working proof-of-concept (PoC) or it does not. This coarse granularity discards partial progress; an attempt that triggers the vulnerability but cannot achieve full exploitation is recorded identically to zero progress. Binary outcomes also produce no signal for two adjacent research communities, since vulnerability prioritization lacks ground-truth depth data to calibrate scoring models and detection engineering lacks behavioral traces from which to derive signatures.

A review of the literature across PoC generation, vulnerability prioritization, and detection rule engineering reveals minimal intersection connecting these domains. PoC generation systems do not produce prioritization signals. Prioritization frameworks do not validate predictions against exploitation data. Detection rule generators, whether classical (TaintCheck~\cite{newsome2005taintcheck}, COVERS~\cite{liang2005covers}) or LLM-based (FALCON~\cite{mitra2025falcon}, RuleGenie~\cite{shukla2025rulegenie}), derive signatures from vulnerability descriptions or malware samples, never from dynamic exploitation telemetry.

A further limitation constrains all benchmark-driven approaches, namely the dependence on pre-built containers. CVE-Bench curates 40 critical-severity web-application CVEs~\cite{zhu2025cvebenchagents} and LiveCVEBench provides 190 tasks across 14 languages and 153 repositories~\cite{cvefactory2026}. These require manual effort per CVE and exhibit high failure rates when configuration drift or port conflicts arise.

To address these gaps, FORGE is presented, a multi-agent system for graduated vulnerability exploitability assessment with three contributions.

\begin{itemize}[leftmargin=1.5em,nosep]
\item \textbf{Graduated exploitation taxonomy.} A four-level depth taxonomy (L0--L3) replaces binary pass/fail with measurable exploitation depth. An LLM-primary oracle evaluates every tool call against CWE-specific criteria, application source code, and server-side snapshots, producing structured verdicts with confidence scores. Structural caps enforce conservative bounds for CWE types where full compromise is architecturally impossible.

\item \textbf{Three-silo bridge.} A single pipeline produces three outputs from each CVE assessment: (1)~graduated exploitation depth as a pattern-level reachability signal for EPSS/CVSS validation, (2)~Sigma and Snort detection rules generated from OpenTelemetry exploitation traces, and (3)~per-CWE attack knowledge that accumulates across assessments. Graduated depth is the bridging mechanism: L2+ exploitation produces behavioral signals rich enough for high-precision detection rules, while L1+ data provides ground truth for prioritization validation.

\item \textbf{Targeted app generation with knowledge transfer.} A Generator Agent creates targeted vulnerable web applications from CVE metadata, eliminating benchmark container dependency. Five knowledge stores, a four-tier intelligence store, generation recipes, auto-distilled cookbook tips, package-specific experience stores, and a per-CWE detection knowledge base, accumulate across assessments, with measurable benefits for generation reliability of known CWE types and exploit efficiency for specific vulnerability classes.
\end{itemize}

Evaluation on 603 CVEs sourced from the CVE-GENIE 841-entry dataset~\cite{ullah2025cvegenie} reaches L1+ on 409/603 (67.8\%) end-to-end at \$1.50 per CVE; 424/603 (70.3\%) pass generation and deployment, of which 96.5\% reach L1+. The system spans eight programming languages (Python, PHP, JavaScript/TypeScript, Go, Ruby, Java, C, and others), 187 unique CWE types, and 440 distinct GitHub repositories.

The remainder of this paper is organized as follows. Section~\ref{sec:background} surveys related work across the three research silos. Section~\ref{sec:architecture} presents the methodology. Section~\ref{sec:evaluation} describes the experimental protocol and results. Section~\ref{sec:discussion} addresses ethical considerations, limitations, and implications. Section~\ref{sec:conclusion} concludes.

\section{Background and Related Work}
\label{sec:background}

Automated vulnerability exploitability assessment draws on three distinct research communities that, despite addressing adjacent problems, have minimal methodological overlap.

\subsection{Automated PoC and Exploit Generation}
\label{sec:bg:poc}

LLM-based vulnerability reproduction has progressed rapidly since 2024, building on the ReAct paradigm~\cite{yao2023react} that interleaves LLM reasoning with tool-use actions. Fang et al.~\cite{fang2024exploit} demonstrate that GPT-4 agents can exploit 87\% of 15 one-day CVEs given their descriptions; Deng et al.~\cite{deng2024pentestgpt} introduce PentestGPT for interactive penetration testing guided by LLMs; Zhu et al.~\cite{zhu2024hptsa} extend this with HPTSA, a hierarchical planning agent that spawns exploitation subagents, achieving 4.3$\times$ improvement over single agents on zero-day targets. CVE-GENIE~\cite{ullah2025cvegenie}, the closest comparable system, reproduces CVEs by cloning the \emph{original} vulnerable project inside a VM and verifying a CTF-style flag, achieving 51\% binary reproduction across 841 CVEs (22 languages) at \$2.77 per CVE. CVE-Bench~\cite{zhu2025cvebenchagents} curates 40 critical-severity web-application CVEs and reports 13\% best-case success under a hierarchical multi-agent framework in the one-day setting. PoCGen~\cite{simsek2025pocgen} targets npm packages specifically, combining taint-path analysis with LLM generation to achieve 77\% success on a curated subset. DrillAgent~\cite{li2026drillagent} introduces execution-state-aware proof-of-vulnerability generation. GONDAR~\cite{gondar2026} employs CWE-specific sink API catalogs with separate exploration and exploitation agents. Liu et al.~\cite{liu2025webvuln} provide the first comprehensive evaluation of 20 LLM agents for web vulnerability reproduction across 16 dimensions. ZeroDayBench~\cite{zerodaybench2026} evaluates frontier agents on 22 novel zero-day vulnerabilities, reporting that even the best models find and patch only a fraction.

All these systems share two fundamental limitations. First, binary evaluation: a CVE either has a working PoC or it does not. Partial progress is not evaluated, preventing correlation with severity metrics and losing information useful for detection engineering. Second, original-project reproduction: CVE-GENIE, CVE-Bench, and PoCGen all require access to the original vulnerable codebase. When complex build dependencies or infrastructure requirements resist automated provisioning, reproduction fails entirely; CVE-GENIE reports 110 setup failures, 198 time/cost-limit failures, and 52 other VM or LLM errors out of 841 candidates~\cite{ullah2025cvegenie}. A fundamentally different approach is taken in FORGE: a targeted minimal application is generated that imports the \emph{actual} vulnerable package at its vulnerable version, exercising the real vulnerability in a simplified environment with infrastructure designed to enable exploitation at each depth level (L1--L3).

Elder et al.~\cite{elder2024survey} systematize 39 exploitability assessment approaches; Bui et al.~\cite{bui2025slr} survey automated exploit and security test generation more broadly. No surveyed approach employs graduated depth measurement.

\subsection{Vulnerability Prioritization}
\label{sec:bg:prioritization}

Organizations triage thousands of CVEs using two primary scoring systems. EPSS~\cite{jacobs2020epss} predicts 30-day exploitation probability from metadata features (CVE age, vendor, CWE type, reference count, partner data). EPSSv3~\cite{jacobs2023epssv3} improves prediction accuracy by 82\% over v2. CVSS provides severity scores based on vulnerability characteristics (attack vector, complexity, impact). SSVC~\cite{ssvc2026} offers stakeholder-specific categorization but requires manual assessment of its Exploitation decision point (none/PoC/active). NIST's LEV framework~\cite{nist2025lev} similarly highlights the absence of automated exploitation evidence in prioritization.

Neither system is validated against actual exploitation attempts. EPSS uses exploit-code existence as a primary feature but never verifies whether that code actually achieves exploitation (beyond the information provided through partner data). Koscinski et al.~\cite{koscinski2025conflicting} demonstrate significant disagreement between CVSS, SSVC, and EPSS on identical CVEs. Parla~\cite{parla2024epss} shows EPSS scores are frequently low or unavailable when high-severity vulnerabilities enter CISA's Known Exploited Vulnerabilities catalog. Gordeychik~\cite{gordeychik2025epss} confirms this pattern with longitudinal analysis of EPSS at KEV addition time. Al~Haddad et al.~\cite{alhaddad2025ssvc} evaluate LLMs for SSVC decision-point prediction, finding consistent over-prediction and poor agreement with expert labels, reinforcing that metadata-based prioritization remains unreliable without ground-truth exploitation data.

No automated system feeds, at scale, exploitation results back to calibrate these prioritization models. In FORGE, ground-truth exploitation depth data are produced that can be directly correlated with EPSS and CVSS scores.

\subsection{Detection Rule Generation}
\label{sec:bg:detection}

\begin{sloppypar}
Detection signature generation from exploit execution has a long history. TaintCheck~\cite{newsome2005taintcheck} generates network signatures via taint tracking during exploit execution. Brumley et al.~\cite{brumley2006signatures} produce vulnerability signatures through symbolic execution. Kobayashi et al.~\cite{kobayashi2023ids} represent the most recent classical approach, generating IDS signatures by executing PoC code and exploring execution paths.
\end{sloppypar}

Between these early systems and the LLM era, detection rule generation from dynamic exploitation traces was largely dormant as a research direction; the community focused instead on signature derivation from static vulnerability descriptions and malware samples. LLM-based detection rule generation has revived the area from a different angle. FALCON~\cite{mitra2025falcon} generates Snort and YARA rules from CTI reports. EvoSIEM~\cite{vy2025evosiem} uses LLMs to enhance Sigma rules against evasion, motivated by evidence that deployed SIEM rules are frequently bypassable~\cite{uetz2024sigmaevasion}. Fairbanks and Serra~\cite{fairbanks2025sigma} automate TTP extraction and Sigma synthesis from CTI text. All generate rules from \emph{descriptions} of attacks (vulnerability advisories, CTI reports, malware samples), not from observed exploitation behavior.

\begin{sloppypar}
Breach and Attack Simulation (BAS) platforms occupy the adjacent space. CALDERA~\cite{applebaum2016caldera} automates post-compromise attack execution using ATT\&CK techniques; S\'{a}nchez-Matas et al.~\cite{sanchezmatas2025bas} combine chaos engineering with CALDERA. These platforms execute attacks but produce no detection content.
\end{sloppypar}

This gap motivates the present work. No existing system generates detection rules from dynamic exploitation telemetry. FORGE's Detector Agent receives OpenTelemetry spans covering every HTTP request, command execution, file access, and server response captured during exploitation, and produces Sigma/Snort rules validated against those same spans.

Two key differentiators distinguish FORGE from these systems. First, graduated depth measurement (L0--L3) replaces binary pass/fail, enabling prioritization ground truth and exploitation-grounded detection rules. Second, targeted environment generation decouples assessment from source-code availability: CVE-GENIE requires the original project, CVE-Bench uses curated containers, and PoCGen targets npm packages only; FORGE generates minimal applications that import the actual vulnerable package, supporting eight languages. No existing system combines exploitation, prioritization validation, and detection rule generation in a single pipeline with cross-CVE knowledge transfer.

\section{Methodology}
\label{sec:architecture}

\begin{figure}[t]
\centering
\begin{tikzpicture}[
  x=1cm, y=1cm,
  arr/.style={-{Stealth[length=3pt, width=2pt]}, gray!60!black, line width=0.4pt},
  oarr/.style={{Stealth[length=2.5pt, width=2pt]}-{Stealth[length=2.5pt, width=2pt]},
    orange!60!black, line width=0.4pt, dashed},
  karr/.style={-{Stealth[length=2pt, width=1.5pt]}, violet!50, line width=0.3pt, dashed},
  pbox/.style={draw=#1!35, fill=#1!3, rounded corners=3pt,
    minimum width=3.5cm, minimum height=1.85cm,
    anchor=north west, line width=0.4pt},
  pbox/.default=blue,
  kbox/.style={rectangle, rounded corners=2pt, draw=violet!30, fill=violet!4,
    inner xsep=2pt, inner ysep=1.5pt, font=\tiny, align=center,
    line width=0.35pt, minimum width=1.85cm},
  lbl/.style={font=\fontsize{5pt}{6pt}\selectfont\itshape, color=gray!60!black},
]

\node[rectangle, rounded corners=3pt, draw=gray!50, fill=gray!5,
  inner xsep=4pt, inner ysep=2pt, font=\scriptsize, line width=0.4pt]
  (input) at (1.75, 0.5)
  {\textbf{CVE-2025-30370} {\tiny CWE-78 $\cdot$ jupyterlab-git $\cdot$ Python}};

\node[pbox=blue] (p1) at (0, 0) {};
\fill[blue!70!black, rounded corners=2pt]
  ([shift={(0.5pt,-0.5pt)}]p1.north west)
  rectangle ([shift={(-0.5pt,-0.35cm)}]p1.north east);
\node[circle, fill=white, text=blue!70!black,
  font=\tiny\bfseries, minimum size=0.28cm, inner sep=0pt]
  at ([shift={(0.3cm,-0.175cm)}]p1.north west) {1};
\node[font=\scriptsize\bfseries, text=white, anchor=west]
  at ([shift={(0.55cm,-0.175cm)}]p1.north west) {Intel Agent};
\node[font=\tiny\itshape, text=gray!60!black, anchor=north west]
  at ([shift={(0.12cm,-0.45cm)}]p1.north west) {10 turns $\cdot$ 6 tools};
\node[font=\tiny, text width=3.2cm, anchor=north west]
  at ([shift={(0.12cm,-0.65cm)}]p1.north west)
  {NVD, OSV, ExploitDB, patches\\PackageResolver: registry validation};
\node[font=\tiny, text=green!40!black, text width=3.2cm, anchor=north west]
  at ([shift={(0.12cm,-1.5cm)}]p1.north west)
  {$\triangleright$ \texttt{jupyterlab-git 0.51, Tier\,2}};

\node[pbox=purple] (p2) at (3.7, 0) {};
\fill[purple!70!black, rounded corners=2pt]
  ([shift={(0.5pt,-0.5pt)}]p2.north west)
  rectangle ([shift={(-0.5pt,-0.35cm)}]p2.north east);
\node[circle, fill=white, text=purple!70!black,
  font=\tiny\bfseries, minimum size=0.28cm, inner sep=0pt]
  at ([shift={(0.3cm,-0.175cm)}]p2.north west) {2};
\node[font=\scriptsize\bfseries, text=white, anchor=west]
  at ([shift={(0.55cm,-0.175cm)}]p2.north west) {Generator};
\node[font=\tiny\itshape, text=gray!60!black, anchor=north west]
  at ([shift={(0.12cm,-0.45cm)}]p2.north west) {5 turns $\cdot$ 4 tools $\cdot$ 2 retries};
\node[font=\tiny, text width=3.2cm, anchor=north west]
  at ([shift={(0.12cm,-0.65cm)}]p2.north west)
  {Actor-critic loop, AppVerifier\\Anti-fabrication, CWE recipes};
\node[font=\tiny, text=green!40!black, text width=3.2cm, anchor=north west]
  at ([shift={(0.12cm,-1.2cm)}]p2.north west)
  {$\triangleright$ \texttt{2 att, Python/pip}};

\node[pbox=teal] (p3) at (7.2, 0) {};
\fill[teal!80!black, rounded corners=2pt]
  ([shift={(0.5pt,-0.5pt)}]p3.north west)
  rectangle ([shift={(-0.5pt,-0.35cm)}]p3.north east);
\node[circle, fill=white, text=teal!80!black,
  font=\tiny\bfseries, minimum size=0.28cm, inner sep=0pt]
  at ([shift={(0.3cm,-0.175cm)}]p3.north west) {3};
\node[font=\scriptsize\bfseries, text=white, anchor=west]
  at ([shift={(0.55cm,-0.175cm)}]p3.north west) {Deploy};
\node[font=\tiny\itshape, text=gray!60!black, anchor=north west]
  at ([shift={(0.12cm,-0.45cm)}]p3.north west) {Podman $\cdot$ resource limits};
\node[font=\tiny, text width=3.2cm, anchor=north west]
  at ([shift={(0.12cm,-0.65cm)}]p3.north west)
  {Dockerfile build $\to$ health check\\Network isolation, port mapping};
\node[font=\tiny, text=green!40!black, text width=3.2cm, anchor=north west]
  at ([shift={(0.12cm,-1.5cm)}]p3.north west)
  {$\triangleright$ \texttt{python:3.11, :8080} {\color{green!50!black}\checkmark}};

\node[pbox=orange] (p4) at (0, -2.7) {};
\fill[orange!70!black, rounded corners=2pt]
  ([shift={(0.5pt,-0.5pt)}]p4.north west)
  rectangle ([shift={(-0.5pt,-0.35cm)}]p4.north east);
\node[circle, fill=white, text=orange!70!black,
  font=\tiny\bfseries, minimum size=0.28cm, inner sep=0pt]
  at ([shift={(0.3cm,-0.175cm)}]p4.north west) {4};
\node[font=\scriptsize\bfseries, text=white, anchor=west]
  at ([shift={(0.55cm,-0.175cm)}]p4.north west) {Planner};
\node[font=\tiny\itshape, text=gray!60!black, anchor=north west]
  at ([shift={(0.12cm,-0.45cm)}]p4.north west) {5 turns $\cdot$ pure reasoning};
\node[font=\tiny, text width=3.2cm, anchor=north west]
  at ([shift={(0.12cm,-0.65cm)}]p4.north west)
  {CWE-aware attack planning\\Primary + escalation + fallback};
\node[font=\tiny, text=green!40!black, text width=3.2cm, anchor=north west]
  at ([shift={(0.12cm,-1.3cm)}]p4.north west)
  {$\triangleright$ \texttt{shell-escape \$()\ $\to$ RCE}};

\node[pbox=red] (p5) at (3.7, -2.7) {};
\fill[red!70!black, rounded corners=2pt]
  ([shift={(0.5pt,-0.5pt)}]p5.north west)
  rectangle ([shift={(-0.5pt,-0.35cm)}]p5.north east);
\node[circle, fill=white, text=red!70!black,
  font=\tiny\bfseries, minimum size=0.28cm, inner sep=0pt]
  at ([shift={(0.3cm,-0.175cm)}]p5.north west) {5};
\node[font=\scriptsize\bfseries, text=white, anchor=west]
  at ([shift={(0.55cm,-0.175cm)}]p5.north west) {Exploit};
\node[font=\tiny\itshape, text=gray!60!black, anchor=north west]
  at ([shift={(0.12cm,-0.45cm)}]p5.north west) {20 turns $\cdot$ 5 tools $\cdot$ gated};
\node[font=\tiny, text width=3.2cm, anchor=north west]
  at ([shift={(0.12cm,-0.65cm)}]p5.north west)
  {T1--3 recon only, T4+ exploit\\L0$\to$L1$\to$L2$\to$L3 progression};
\node[font=\tiny, text=green!40!black, text width=3.2cm, anchor=north west]
  at ([shift={(0.12cm,-1.2cm)}]p5.north west)
  {$\triangleright$ {\color{red!70!black}\texttt{4T, L1$\to$L3 via \$()}}};

\node[pbox=indigo] (p6) at (7.2, -2.7) {};
\fill[indigo, rounded corners=2pt]
  ([shift={(0.5pt,-0.5pt)}]p6.north west)
  rectangle ([shift={(-0.5pt,-0.35cm)}]p6.north east);
\node[circle, fill=white, text=indigo,
  font=\tiny\bfseries, minimum size=0.28cm, inner sep=0pt]
  at ([shift={(0.3cm,-0.175cm)}]p6.north west) {6};
\node[font=\scriptsize\bfseries, text=white, anchor=west]
  at ([shift={(0.55cm,-0.175cm)}]p6.north west) {Detector};
\node[font=\tiny\itshape, text=gray!60!black, anchor=north west]
  at ([shift={(0.12cm,-0.45cm)}]p6.north west) {15 turns $\cdot$ if L1+ $\cdot$ OTEL};
\node[font=\tiny, text width=3.2cm, anchor=north west]
  at ([shift={(0.12cm,-0.65cm)}]p6.north west)
  {Detection KB (10 CWEs $\times$ 4 tiers)\\Sigma + Snort rule generation};
\node[font=\tiny, text=green!40!black, text width=3.2cm, anchor=north west]
  at ([shift={(0.12cm,-1.5cm)}]p6.north west)
  {$\triangleright$ \texttt{2 rules, sigma + snort}};

\draw[arr] (input.south) -- (p1.north);

\draw[arr] ([yshift=-0.9cm]p1.north east) -- ([yshift=-0.9cm]p2.north west);
\draw[arr] ([yshift=-0.9cm]p2.north east) -- ([yshift=-0.9cm]p3.north west);

\draw[arr, rounded corners=3pt]
  (p3.south) -- ++(0,-0.35) -| (p4.north);

\draw[arr] ([yshift=-0.9cm]p4.north east) -- ([yshift=-0.9cm]p5.north west);
\draw[arr] ([yshift=-0.9cm]p5.north east) -- ([yshift=-0.9cm]p6.north west);

\node[rectangle, rounded corners=2pt, draw=red!40, fill=red!5,
  inner xsep=3pt, inner ysep=2pt, font=\tiny, line width=0.4pt,
  anchor=north] (out) at ([yshift=-0.15cm]p6.south)
  {{\scriptsize\bfseries\color{red!70!black} L3} {\tiny\color{gray!60!black}$\cdot$ \$0.95 $\cdot$ 4m\,54s}};
\draw[arr] (p6.south) -- (out.north);

\node[rectangle, rounded corners=3pt, draw=orange!40, fill=orange!4,
  inner sep=3pt, text width=4.2cm, font=\tiny, line width=0.4pt,
  anchor=north] (oracle) at (5.47, -4.8)
  {{\scriptsize\bfseries\color{orange!70!black} LLM-Primary Oracle}\\[1pt]
   Per-turn: tool req/resp + CWE criteria + app source\\[1pt]
   Caps: XSS,CSRF $\to$ L2; coaching hints};

\draw[oarr] (p5.south) -- (oracle.north)
  node[lbl, midway, right, xshift=1pt] {every turn};

\node[kbox] (k1) at (1.0, -6.8) {\textbf{Intel Store}\\4 tiers, 90d decay};
\node[kbox] (k2) at (3.1, -6.8) {\textbf{Gen.\ KB}\\CWE recipes};
\node[kbox] (k3) at (5.2, -6.8) {\textbf{Cookbook}\\Build tips/lang};
\node[kbox] (k4) at (7.3, -6.8) {\textbf{Pkg Exp.}\\Episodic+semantic};
\node[kbox] (k5) at (9.6, -6.8) {\textbf{Det.\ KB}\\10 CWEs $\times$ 4 tiers};

\begin{scope}[on background layer]
  \node[rectangle, rounded corners=3pt, draw=violet!30, fill=violet!2,
    line width=0.3pt, fit=(k1)(k2)(k3)(k4)(k5),
    inner xsep=6pt, inner ysep=3pt,
    yshift=0pt] (kbg) {};
  \path let \p1=(kbg.north west), \p2=(kbg.north east) in
    node[rectangle, rounded corners=3pt, draw=violet!30, fill=violet!2,
      line width=0.3pt, anchor=south west,
      minimum width={\x2-\x1}, minimum height=0.35cm]
      (kbanner_bg) at (\p1) {};
\end{scope}

\begin{scope}[on background layer]
  \node[rectangle, rounded corners=3pt, draw=violet!30, fill=violet!2,
    line width=0.3pt, fit=(kbg)(kbanner_bg),
    inner sep=0pt] (kfull) {};
\end{scope}

\fill[violet!60!black, rounded corners=2pt]
  ([shift={(0.5pt,-0.5pt)}]kfull.north west)
  rectangle ([shift={(-0.5pt,-0.35cm)}]kfull.north east);
\node[font=\scriptsize\bfseries, text=white, anchor=west]
  at ([shift={(0.15cm,-0.175cm)}]kfull.north west) {Knowledge Architecture};

\end{tikzpicture}
\caption{FORGE pipeline architecture with running example (CVE-2025-30370, jupyterlab-git command-injection RCE via \texttt{\$()} substitution that bypasses partial shell-escaping; L3 in 4 exploit turns). The LLM-primary oracle evaluates every exploit turn against CWE-specific criteria. Five knowledge stores accumulate intelligence across assessments.}
\label{fig:architecture}
\end{figure}
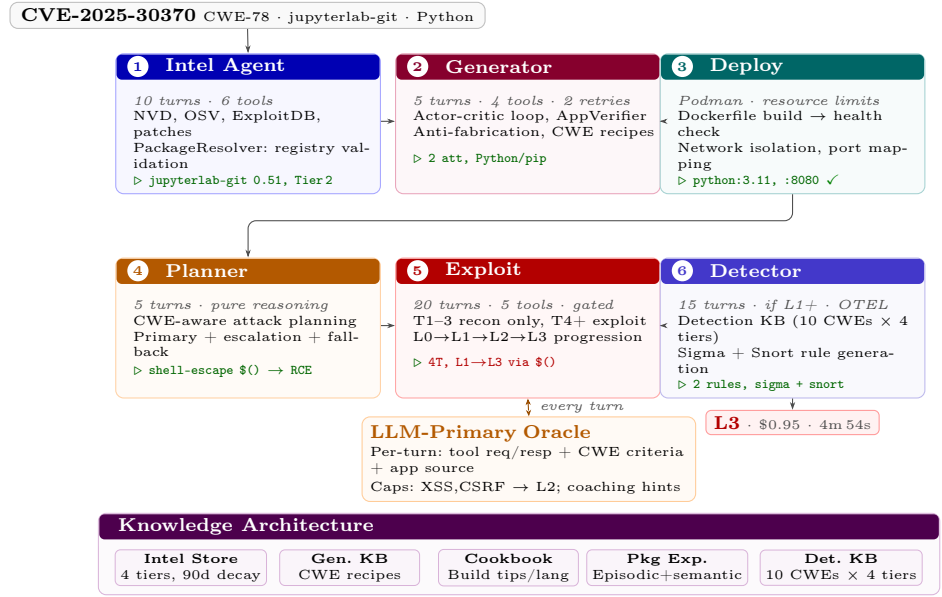

FORGE comprises five specialized agents executing in a fixed pipeline, a tiered oracle for per-turn exploitation assessment, and a knowledge architecture that accumulates intelligence across CVE assessments. Figure~\ref{fig:architecture} illustrates the pipeline topology.

\subsection{Pipeline Overview}
\label{sec:arch:pipeline}

The pipeline processes a single CVE through five stages. Each agent operates within a fixed turn budget and restricted toolset, enforcing bounded cost per assessment.

The \textbf{Intel Agent} compiles a structured intelligence report containing root cause analysis, attack surface characterization, and technology stack identification from NVD advisories, GitHub security advisories, patch diffs, and known exploit databases.

A \textbf{PackageResolver} then resolves the vulnerable package name, version, and ecosystem from NVD, OSV, and CVE-GENIE metadata. Candidate versions are validated against actual package registries (npm, PyPI, RubyGems, Packagist); if the metadata version does not exist in the registry, the resolver falls back to the closest available version. This step prevents generation failures caused by stale or non-existent version strings in vulnerability databases.

The \textbf{Generator Agent} creates a targeted vulnerable web application that embeds the \emph{actual} vulnerable package at its vulnerable version. Unlike systems that clone the original project (e.g., CVE-GENIE~\cite{ullah2025cvegenie}), a minimal application is generated with infrastructure explicitly shaped by the exploitation taxonomy. The key design decision is \emph{level-aware scaffolding}: for L1, vulnerable endpoints propagate exceptions to HTTP responses rather than suppressing errors; for L2, real data infrastructure (SQLite databases seeded with user records, sensitive configuration files) enables meaningful data exfiltration evidence; for L3, code execution paths remain unsandboxed and writable directories exist for marker file creation. This ensures the exploit agent encounters a surface that can produce graduated evidence, not just a binary trigger.

Input includes the intelligence report, CWE-specific generation recipes, resolved package information, and package-specific build experience from prior assessments. An actor-critic validation loop builds, deploys, and tests each generated application; on failure, structured feedback drives the next attempt. An \texttt{AppVerifier} performs multi-layer fabrication detection: verifying that declared dependencies are actually imported in source code, scanning for mock-service patterns (e.g., \texttt{FakeRedis}, raw TCP socket servers), and performing runtime port scans inside the container to detect fake service listeners.

The \textbf{Planner Agent}, a pure reasoning agent with no tools, produces a ranked attack plan with escalation paths from the intelligence report, application manifest, and package-specific exploitation experience.

\begin{sloppypar}
The \textbf{Exploit Agent} executes the attack plan against the deployed application. Tool gating is the critical design choice: the first three LLM calls restrict the agent to reconnaissance-only tools, preventing premature exploitation attempts before the attack surface is understood. Three calls was set empirically to prevent exploitation before vulnerability context is processed; all tools are available from LLM call~4. The LLM-primary oracle evaluates after every tool call, and coaching messages are injected when the agent fails to progress. A configurable grace period (3~turns after L3, 5~turns after L2) allows evidence consolidation before forced exit.
\end{sloppypar}

Finally, the \textbf{Detector Agent} runs only if exploitation reaches L1 or higher. It queries OpenTelemetry spans from the exploitation phase, retrieves reference rules from a curated detection knowledge base, and generates Sigma and Snort rules grounded in the collected exploitation signals.

\subsection{Graduated Exploitation Taxonomy}
\label{sec:arch:taxonomy}

Table~\ref{tab:taxonomy} defines the four exploitation levels. The running maximum never decreases during a single assessment; once a level is reached, it persists regardless of subsequent turn outcomes.

\begin{table}[t]
\caption{Graduated exploitation taxonomy (L0--L3).}\label{tab:taxonomy}
\centering
\small
\setlength{\tabcolsep}{4pt}
\begin{tabular}{@{}clp{4.7cm}@{}}
\toprule
\textbf{Level} & \textbf{Name} & \textbf{Evidence} \\
\midrule
L0 & No evidence & No exploitation evidence observed \\
L1 & Triggered & Targeted payload activates the specific vulnerable functionality: Path traversal sequence returns file contents, XSS payload reflected without encoding \\
L2 & Exploited & Significant attacker-controlled impact: data exfiltration, code execution evidence, arbitrary file access \\
L3 & Compromised & Full compromise: RCE with output, complete data dump, persistence, privilege escalation \\
\bottomrule
\end{tabular}
\end{table}

\subsection{LLM-Primary Oracle}
\label{sec:arch:oracle}

Central to the graduated assessment is the oracle, which determines exploitation level; agent self-assessment is ignored. For every tool call, the oracle receives: (1)~the tool request and response, (2)~CWE-specific evidence criteria covering 20 CWE groups with natural-language descriptions for each level, (3)~the generated application's source code, and (4)~a server-side snapshot (created files, running processes, log tail from the sandbox).

The oracle produces a structured verdict: level (0--3), confidence (0.0--1.0), and reasoning. Structural caps enforce conservative bounds for CWE types where full compromise is architecturally impossible: XSS (CWE-79), path traversal (CWE-22), CSRF (CWE-352), and denial-of-service (CWE-400) are capped at L2 maximum. Tool-level caps prevent inflation from sandbox-local operations: \texttt{read\_file} is capped at L1 unless the file was created by the exploit; \texttt{exec\_command} is capped at L1 unless the command targets the application.

This design addresses the reward-hacking failure mode documented in the LLM-as-judge literature~\cite{denison2024sycophancy}: the exploit agent cannot inflate its own score through self-congratulatory narrative because the oracle evaluates tool outputs, not agent text.

\subsection{Knowledge Architecture}
\label{sec:arch:knowledge}

Underpinning the entire pipeline, five knowledge stores accumulate across assessments:

\paragraph{Four-Tier Intelligence Store.} Tier~1 caches raw API responses. Tier~2 stores compiled per-CVE intelligence reports. Tier~3 accumulates per-CWE attack knowledge with 90-day confidence decay, strengthened by confirmations. Tier~4 is an append-only procedural memory of cross-CVE exploitation patterns.

\paragraph{Generation Knowledge Base.} Per-CWE design recipes that record framework-specific generation patterns with graduated complexity levels. Success rates are updated after each generation attempt.

\paragraph{Cookbook Store.} Auto-distilled tips extracted by an LLM critic from build errors and successful exploitation patterns. Organized by language and CWE type, with a 20-tip cap per category to prevent unbounded growth.

\paragraph{Package Experience Store.} This store accumulates package-specific learnings keyed by package name and ecosystem rather than CWE type. Three memory tiers, episodic (per-CVE history), semantic (LLM-distilled patterns), and procedural (concrete recipes), capture that building and exploiting a given package has consistent patterns regardless of which CWE is involved. In the CVE-GENIE dataset, 536 of 841 CVEs share a package with at least one other CVE, making cross-CVE knowledge transfer within package families a potential efficiency lever.

\paragraph{Detection Knowledge Base.} Per-CWE reference rules organized across four detection tiers (web, host, network, application). After each assessment, validated Sigma and Snort rules are appended alongside curated exemplars; the Detector Agent for subsequent CVEs of the same CWE retrieves these accumulated rules as generation templates.

\section{Evaluation}
\label{sec:evaluation}

FORGE is evaluated along four dimensions that correspond to the three-silo bridge articulated in Section~\ref{sec:introduction}: exploitation capability (\S\ref{sec:eval:rq1}), correlation with prioritization metrics (\S\ref{sec:eval:rq2}), detection rule structural quality (\S\ref{sec:eval:rq3}), and knowledge transfer efficiency (\S\ref{sec:eval:rq4}).

\subsection{Experimental Setup}
\label{sec:eval:setup}

\paragraph{Dataset.} All CVEs are drawn from the 841-entry CVE-GENIE dataset~\cite{ullah2025cvegenie}. After excluding 37 native-only C/C++ memory safety CVEs and 11 lacking CWE tags, an eligible pool of 793 CVEs remains; 603 are selected via stratified diversity sampling across CWE type, language, and project type. Native memory-corruption CVEs lie outside the targeted-application generation model and are out of scope. The dataset spans 440 GitHub repositories, 187 CWE types, and eight languages. The PackageResolver uses NVD and OSV as primary sources, with CVE-GENIE metadata supplementing resolution when available; performance on CVEs outside this distribution is not measured.

\paragraph{Infrastructure and budget.} Experiments execute on a \texttt{t3.xlarge} AWS EC2 instance (4 vCPU, 16\,GB RAM) with Podman sandboxing. LLM inference uses Claude Sonnet 4.5 (agents) and GPT-5 Mini (oracle critic) via LiteLLM with OpenTelemetry token tracking. Additionally, per-CVE cost is capped at \$2.50.

\paragraph{Oracle validation.} A structured audit is applied at each level: CWE-class evidence rules are matched against the exploit agent's tool-result corpus and oracle summary, yielding VALID (level reach plus evidence in both), BORDERLINE (partial), or INVALID. Across 91 stratified assessments at L1, L2, and L3 (seed=42), 92.3\% are VALID (95\% CI [84.8\%, 96.9\%]); per-level VALID rates fall between 86.7\% and 96.8\%. One L3 case is a likely false positive (authorization bypass reaching L2 evidence but not full compromise); the remaining six non-VALID assessments are BORDERLINE. The audit script and per-CVE verdicts are released with the artifact.

\paragraph{Baseline.} CVE-GENIE~\cite{ullah2025cvegenie} is the closest comparable system (50.9\% binary success at \$2.77/CVE). The two setups occupy different points on the cost-fidelity curve and are not directly comparable: CVE-GENIE reconstructs the original vulnerable project from its source tree, whereas FORGE consumes only CVE metadata and the resolved package coordinates and exploits a minimal downstream integration that imports the vulnerable package. The binary CTF-style flag check is replaced by a graduated L0--L3 verdict, and Sigma/Snort detection rules are produced alongside exploitation. Per-CVE cost is 46\% lower (\$1.50 vs.\ \$2.77). A focused head-to-head against PoCGen~\cite{simsek2025pocgen} on its npm subset is left to future work.

\begin{sloppypar}
\paragraph{Overlap with CVE-GENIE.} Of the 429 CVEs that CVE-GENIE successfully reproduced, 316 appear in the FORGE 603-CVE subset. On this shared set, FORGE reaches L1+ on 227/316 (71.8\%); generation fails on 79 (25.0\%), predominantly Dockerfile build errors. On the 287 CVEs where CVE-GENIE failed, FORGE reaches L1+ on 182 (63.4\%), recovering CVEs whose original-project builds resisted automated provisioning. FORGE-only successes (30.2\% of the 603) and GENIE-only successes (14.8\%) have minimal overlap (OR${=}$1.47, $p{=}$0.029), indicating that the targeted-app and original-project approaches fail on different CVE subsets.
\end{sloppypar}

\subsection{Exploitation Depth at Scale}
\label{sec:eval:rq1}

\begin{table}[t]
\caption{Exploitation depth across 603 CVEs.}\label{tab:rq1}
\centering
\small
\begin{tabular}{@{}lr@{}}
\toprule
\textbf{Metric} & \textbf{Value} \\
\midrule
CVEs attempted & 603 \\
Generation+deployment success & 424/603 (70.3\%) \\
L1+ rate (end-to-end) & 409/603 (67.8\%) \\
L1+ rate (of generated) & 409/424 (96.5\%) \\
Mean exploitation level (L1+) & 1.88 \\
L3 (full compromise) rate & 33/424 (7.8\%) \\
Mean cost per CVE & \$1.50 \\
Mean exploit turns per CVE & 9.5 \\
\midrule
\multicolumn{2}{@{}l}{\textit{Level distribution (of generated):}} \\
\quad L0 (no evidence) & 15 (3.5\%) \\
\quad L1 (triggered) & 81 (19.1\%) \\
\quad L2 (exploited) & 295 (69.6\%) \\
\quad L3 (compromised) & 33 (7.8\%) \\
\bottomrule
\end{tabular}
\end{table}

\begin{figure}[t]
\centering
\subfloat[Level distribution across 603 CVEs.\label{fig:level_dist}]{%
  \includegraphics[width=0.48\columnwidth]{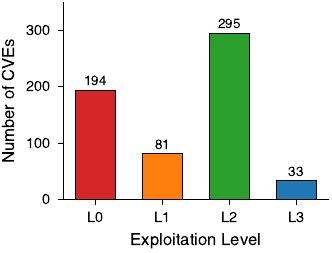}}\hfil
\subfloat[EPSS vs.\ exploitation level.\label{fig:epss_scatter}]{%
  \includegraphics[width=0.48\columnwidth]{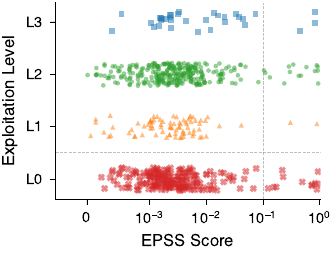}}
\caption{Exploitation depth and prioritization metric correlation (RQ1--RQ2).}\label{fig:rq1_rq2}
\end{figure}

\begin{figure}[t]
\centering
\includegraphics[width=\columnwidth]{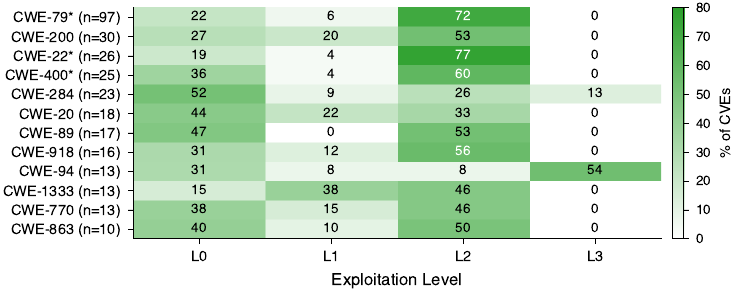}
\caption{Exploitation level distribution for the 12 most frequent primary CWEs (\% of CVEs in each row reaching each level). Cell shade is the within-CWE share of the L0--L3 column; rows marked with~$\ast$ are structurally capped at L2 by the oracle (\S\ref{sec:arch:oracle}).}
\label{fig:cwe_heatmap}
\end{figure}

Of 603 CVEs attempted, 424 (70.3\%) pass generation and deployment, and 409/603 (67.8\%) reach L1+ end-to-end (Table~\ref{tab:rq1}, Figure~\ref{fig:level_dist}). Of the 424 generated applications, 96.5\% reach L1+, concentrated at L2 (69.6\%); 91.9\% of exploited CVEs fall at L1--L2, which a binary metric would collapse into ``success''. Bootstrap 95\% CI for mean depth across all 603 CVEs (coding generation failures as L0) is [1.20, 1.36]. Generation failure (29.7\%) is the primary bottleneck and varies sharply by language: Rust (74\%), TypeScript (45\%), and Ruby (39\%) fail most often, while PHP (6\%), C\# (5\%), and Java (17\%) rarely fail. Figure~\ref{fig:cwe_heatmap} breaks the same data down by primary CWE: capped CWEs (XSS, CWE-22, CWE-352, CWE-400; marked~$\ast$) cluster at L2 by construction, while code-execution and deserialization classes (CWE-94, CWE-502, CWE-78) carry most of the L3 mass. After the L2 caps, 314/424 (74.1\%) generated applications are L3-eligible; the conditional L3 rate among eligible is 33/314 (10.5\%). Per-CVE cost averages \$1.50.

\subsection{Correlation with Prioritization Metrics}
\label{sec:eval:rq2}

Given that a substantial fraction of CVEs can be exploited at graduated depth, a natural question is whether this depth correlates with existing prioritization metrics. FORGE exploitation depth is correlated with EPSS and CVSS scores across all 603 CVEs (Figure~\ref{fig:epss_scatter}). Neither metric exhibits significant rank correlation with exploitation level (Spearman $\rho{=}0.062$, $p{=}0.13$ for EPSS; $\rho{=}0.016$, $p{=}0.71$ for CVSS). Of 409 exploited CVEs, 389 (95.1\%) carry EPSS below 0.1, although low-EPSS CVEs are not \emph{disproportionately} exploitable; rather, 94.7\% of all dataset CVEs fall below this threshold.

The key finding emerges from per-band analysis: CVEs are exploited at ${\sim}68\%$ regardless of EPSS band ($<$0.001: 68.6\%, 0.001--0.01: 68.0\%, 0.01--0.1: 67.7\%, $\geq$0.1: 62.5\%), and an equally flat profile is observed across CVSS severity bands. It is important to note that graduated depth measures a distinct property: how far exploitation proceeds once code-level vulnerability conditions are satisfied. The flat exploitation profile across all scoring bands provides large-scale empirical evidence that metadata-based prioritization scores do not predict code-level exploitation depth. However, the dataset is drawn from a single source (CVE-GENIE) that may be homogeneous in exploitability characteristics; the flat profile could partially reflect dataset composition rather than a universal property of CVEs. That no predictive relationship is found suggests graduated depth constitutes an independent measurement axis, one that automated assessment can now produce at scale for any CVE with a package-registry entry.

\subsection{Detection Rule Structural Quality}
\label{sec:eval:rq3}

\begin{table}[t]
\caption{Detection rule grounding stratified by exploitation depth.}\label{tab:rq3}
\centering
\small
\begin{tabular}{@{}lcc@{}}
\toprule
\textbf{Stratum} & \textbf{Raw} & \textbf{Norm.} \\
\midrule
L1 ($n{=}63$)  & 0.456 & 0.175 \\
L2+ ($n{=}293$) & 0.336 & 0.307 \\
\midrule
$p$ (raw) & \multicolumn{2}{c}{0.997 (n.s.)} \\
$p$ (norm.) & \multicolumn{2}{c}{\textbf{0.035}} \\
\bottomrule
\end{tabular}
\end{table}

Graduated depth also determines detection rule quality. For each L1+ CVE, the Detector Agent generates Sigma and Snort rules from OTEL spans collected during exploitation (Table~\ref{tab:rq3}). \emph{Span grounding} is defined as the fraction of rule field values (URLs, commands, file paths) that appear verbatim in collected spans, indicating rules anchored in observed behavior rather than hallucinated.

Raw grounding rates are confounded by observation volume: L1 CVEs generate 2.4$\times$ more spans (17.0 vs.\ 7.1) driven by longer exploit traces (16.1 vs.\ 7.8 turns), mechanically inflating substring matches. The quantity of interest for any comparison where span counts differ is grounded fields per span, which answers the operational question of how much detection-relevant evidence each collected span carries. Under this normalized metric, L2+ rules achieve marginally significantly higher grounding (0.307 vs.\ 0.175, $p{=}0.035$): each span from deeper exploitation contributes more detection-relevant evidence. Rules are generated for 380 L1+ CVEs (mean 2.0 rules/CVE). Span grounding measures structural quality (whether rule fields are anchored in observed behavior rather than hallucinated) and not detection efficacy. As a direct test of specificity, the 392 generated Snort rules are evaluated against a 1{,}000-request synthetic benign HTTP corpus: 93.4\% produce zero matches, with zero-FP rates of 95.6\% at L1 ($n{=}68$) and 92.9\% at L2+ ($n{=}324$). The remaining FP-prone rules cluster on CSRF and access-control patterns where benign state-changing POSTs are indistinguishable from attacks without business-logic context. FP rates against benign traffic measure specificity; true-positive validation and operational deployment assessment require a replay-based SIEM evaluation.

\begin{figure}[t]
\centering
\subfloat[Span-normalized grounding by depth ($p{=}0.035$).\label{fig:grounding}]{%
  \includegraphics[width=0.48\columnwidth]{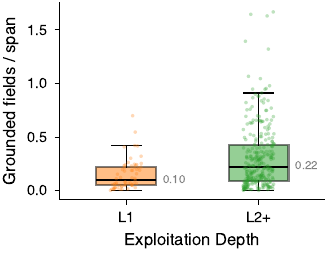}}\hfil
\subfloat[Per-agent cost share.\label{fig:agent_cost}]{%
  \includegraphics[width=0.48\columnwidth]{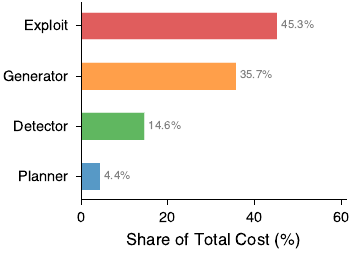}}
\caption{Detection rule quality (RQ3) and cost decomposition (RQ4).}\label{fig:rq3_rq4}
\end{figure}

\subsection{Knowledge Reuse}
\label{sec:eval:rq4}

This dimension asks whether the knowledge architecture leads the LLM to avoid mistakes that earlier CVEs of the same package or CWE already recorded. Generator failure details are parsed from run logs and canonicalized into mistake signatures (wrong package versions, forgotten vulnerable imports, unknown build tags, native-compile failures); mean generation attempts is close to one, so the signature grain replaces cost as the instrument. The exploitation store has accumulated 1{,}758 technique entries across 134 CWE types, the experience store covers 80 multi-CVE packages across 8 ecosystems, and the cookbook store covers 12 languages.
The 79 packages with two or more CVEs yield 250 prior-signature observations across 277 generator logs, of which 48.0\% are avoided on the next CVE for the same package; \texttt{directus/npm}, \texttt{evmos/v18/go}, and \texttt{trix/npm} reach higher levels between 73--83\%. At the CWE grain, 2{,}510 prior exploit dead-end observations across 57 CWEs are avoided 95.5\% of the time on the next CVE of the same CWE (95\% CI [94.7\%, 96.3\%]); exploit-side transfer is markedly stronger than build-side, consistent with the CWE-keyed Tier~3 attack store. Both rates are observational; the KB-injection audit confirms stored signatures were present in the agent prompt at time-of-avoidance, but causal attribution requires a controlled empty-KB ablation, left to future work (\S\ref{sec:disc:limitations}). Per-CVE cost stays stable at \$1.50 across the run with the breakdown shown in Figure~\ref{fig:agent_cost}, and exploitation rates are maintained across 134 CWE types and eight languages without per-CVE cost escalation.

\section{Discussion}
\label{sec:discussion}

\subsection{Ethical Considerations}
\label{sec:disc:ethics}

Targeted applications generated by the system itself are exploited, and no real-world systems are targeted at any point. All exploitation is performed within isolated Podman containers with resource limits and network isolation, and only publicly disclosed CVEs with assigned identifiers are processed.

Nevertheless, functional offensive artifacts (exploit scripts, payload configurations, attack logs) are produced alongside defensive detection rules. Although these target purpose-built applications, they constitute working exploits for the demonstrated vulnerability patterns. Dual-use risk is mitigated by three design choices: (1)~targeted applications lack production defenses (WAFs, authentication), requiring adaptation for real targets; (2)~knowledge stores contain CWE-pattern abstractions, not target-specific chains; (3)~detection rules accompany all exploitation results.

\subsection{Limitations}
\label{sec:disc:limitations}

\paragraph{Application fidelity and scope.} The targeted minimal applications preserve the vulnerability pattern but lack production complexity such as authentication, rate limiting, web application firewalls, and multi-tier architectures; depth therefore reflects pattern-level reachability rather than deployment-specific risk. An L2 result on a FORGE-generated application demonstrates that the vulnerability class permits data exfiltration under favorable conditions, but says nothing about whether the same attack would succeed against a hardened deployment with input validation and network segmentation. Conversely, an L0 result does not imply the vulnerability is unexploitable in production --- it may indicate that the generated application failed to exercise the vulnerable code path, not that the path is unreachable. Results cover 2024--2025 open-source web and library CVEs from a single dataset, and generalization to proprietary software, embedded systems, or multi-component vulnerability classes is not claimed.

\paragraph{Oracle accuracy.} A structured audit across 91 stratified assessments finds 92.3\% VALID overall (95\% CI [84.8\%, 96.9\%]); separate models for agents (Claude Sonnet 4.5) and oracle (GPT-5 Mini) further mitigate feedback loops.

\paragraph{Reproducibility and knowledge attribution.} Commercial LLM APIs introduce non-determinism; temperature is set to~0 for all agents except the Generator (0.2). The avoidance figures are observational across the canonical CVE order; a controlled empty-KB ablation is left to future work.

\subsection{Implications}
\label{sec:disc:implications}

\paragraph{Ground-truth exploitability.} Automated exploitation validates EPSS predictions with behavioral evidence rather than metadata alone. The flat exploitation rate across all EPSS bands implies that metadata-based prioritization and pattern-level reachability measure orthogonal properties. The overlap analysis further indicates that targeted-app and original-project approaches fail on different CVE subsets, suggesting that combining both would maximize coverage.

\paragraph{Detection-capability-centric risk.} Generating detection rules grounded in exploitation behavior transforms the detection gap from binary (rule exists or not) into a quality-graded metric. The evaluation quantifies this: L2+ exploitation produces rules with 75\% higher span-normalized grounding than L1-derived rules, indicating that exploitation depth directly determines detection rule quality and enabling organizations to assess whether their detection infrastructure can observe specific exploitation patterns.

\paragraph{Assessment economics.} At \$1.50 per CVE, continuous vulnerability assessment is made economically viable: against the 48{,}000+ CVEs disclosed annually, full-corpus assessment would cost approximately \$72K/year, although most organizations do not need to evaluate all CVEs. The knowledge architecture maintains consistent exploitation rates across 134 CWE types without per-CVE cost escalation, demonstrating both the increasing capability of LLMs and the efficiencies of the knowledge architecture.

\section{Conclusion}
\label{sec:conclusion}

This paper presented FORGE, a multi-agent system for graduated vulnerability exploitability assessment that bridges three research communities, namely proof-of-concept generation, vulnerability prioritization, and detection engineering. The key technical contributions are a four-level exploitation taxonomy (L0--L3) evaluated by an LLM-primary oracle with CWE-specific structural caps, targeted vulnerable application generation that eliminates benchmark container dependency, and a tiered knowledge architecture that accumulates intelligence across CVE assessments.

Evaluation on 603 CVEs from the CVE-GENIE dataset demonstrates 67.8\% end-to-end L1+ exploitation (409/603) at \$1.50 per CVE across eight languages and 187 CWE types. Graduated depth reveals that exploitation rates remain near 68\% across all EPSS and CVSS bands, indicating that pattern-level reachability is orthogonal to metadata-based prioritization. Detection rules from L2+ exploitation achieve higher span-normalized grounding ($p{=}0.035$), and within-package signature analysis shows 48\% avoidance of previously recorded build-failure modes.

Future work can proceed along three dimensions. First, adaptive pipeline routing (skipping the Planner for well-characterized CWE types) could improve throughput and depth. Second, progressively increasing application complexity beyond minimal reproducers would close the fidelity gap with production environments. Third, longitudinal operation on newly disclosed CVEs would measure knowledge accumulation over months and could feed graduated depth data back to EPSS calibration as a community resource.

All code, per-CVE results, oracle audit verdicts, and analysis scripts are released under an Apache 2.0 license at \url{https://github.com/dynatrace-oss/forge} and \url{https://github.com/dynatrace-oss/forge-artifacts}.

\begin{credits}
\subsubsection{\discintname}
The authors have no competing interests to declare that are relevant to the content of this article.
\end{credits}

\bibliographystyle{splncs04}
\bibliography{references}

@misc{cve2025metrics,
  title={{CVE} Metrics},
  author={{CVE.org}},
  year={2025},
  howpublished={\url{https://www.cve.org/About/Metrics}},
  note={Accessed: February 2026}
}

@article{fang2024exploit,
  title={{LLM} agents can autonomously exploit one-day vulnerabilities},
  author={Fang, Richard and Bindu, Rohan and Gupta, Akul and Kang, Daniel},
  journal={arXiv preprint arXiv:2404.08144},
  year={2024},
  url={https://arxiv.org/abs/2404.08144}
}

@article{zhu2024hptsa,
  title={Teams of {LLM} agents can exploit zero-day vulnerabilities},
  author={Zhu, Yuxuan and Kellermann, Antony and Gupta, Akul and Li, Philip and Fang, Richard and Bindu, Rohan and Kang, Daniel},
  journal={arXiv preprint arXiv:2406.01637},
  year={2024},
  url={https://arxiv.org/abs/2406.01637}
}

@inproceedings{deng2024pentestgpt,
  title={{PentestGPT}: Evaluating and harnessing large language models for automated penetration testing},
  author={Deng, Gelei and Liu, Yi and Mayoral-Vilches, V{\'\i}ctor and Liu, Peng and Li, Yuekang and Xu, Yuan and Zhang, Tianwei and Liu, Yang and Pinzger, Martin and Rass, Stefan},
  booktitle={Proceedings of the 33rd USENIX Security Symposium},
  pages={847--864},
  year={2024},
  url={https://www.usenix.org/conference/usenixsecurity24/presentation/deng}
}

@article{ullah2025cvegenie,
  title={From {CVE} entries to verifiable exploits: An automated multi-agent framework for reproducing {CVEs}},
  author={Ullah, Saad and Balasubramanian, Praneeth and Guo, Wenbo and Burnett, Amanda and Pearce, Hammond and Kruegel, Christopher and Vigna, Giovanni and Stringhini, Gianluca},
  journal={arXiv preprint arXiv:2509.01835},
  year={2025},
  url={https://arxiv.org/abs/2509.01835}
}

@inproceedings{zhu2025cvebenchagents,
  title={{CVE-Bench}: A benchmark for {AI} agents' ability to exploit real-world web application vulnerabilities},
  author={Zhu, Yuxuan and Kellermann, Antony and Bowman, Dylan and Li, Philip and Gupta, Akul and Danda, Adarsh and Fang, Richard and Jensen, Conner and Ihli, Eric and Benn, Jason and Geronimo, Jet and Dhir, Avi and Rao, Sudhit and Yu, Kaicheng and Stone, Twm and Kang, Daniel},
  booktitle={Proceedings of the 42nd International Conference on Machine Learning (ICML)},
  year={2025},
  url={https://arxiv.org/abs/2503.17332}
}

@article{simsek2025pocgen,
  title={{PoCGen}: Generating proof-of-concept exploits for vulnerabilities in npm packages},
  author={Simsek, Deniz and Eghbali, Aryaz and Pradel, Michael},
  journal={arXiv preprint arXiv:2506.04962},
  year={2025},
  url={https://arxiv.org/abs/2506.04962}
}

@article{li2026drillagent,
  title={Execution-state-aware {LLM} reasoning for automated proof-of-vulnerability generation},
  author={Li, Haoyu and Che, Xijia and Wang, Yanhao and Liao, Xiaojing and Xing, Luyi},
  journal={arXiv preprint arXiv:2602.13574},
  year={2026},
  url={https://arxiv.org/abs/2602.13574}
}

@article{gondar2026,
  title={Contextualizing sink knowledge for {Java} vulnerability discovery},
  author={Fleischer, Fabian and Zhang, Cen and Jang, Joonun and Cho, Jeongin and Xu, Meng and Kim, Taesoo},
  journal={arXiv preprint arXiv:2604.01645},
  year={2026},
  url={https://arxiv.org/abs/2604.01645}
}

@article{cvefactory2026,
  title={{CVE-Factory}: Scaling expert-level agentic tasks for code security vulnerability},
  author={Luo, Xianzhen and Zhang, Jingyuan and Zhou, Shiqi and Huang, Rain and Xiao, Chuan and Zhu, Qingfu and Ma, Zhiyuan and Yue, Xing and Yue, Yang and Zeng, Wencong and Che, Wanxiang},
  journal={arXiv preprint arXiv:2602.03012},
  year={2026},
  url={https://arxiv.org/abs/2602.03012}
}

@article{elder2024survey,
  title={A survey on software vulnerability exploitability assessment},
  author={Elder, Sarah and Rahman, Md Rayhanur and Fringer, Gage and Kapoor, Kunal and Williams, Laurie},
  journal={ACM Computing Surveys},
  volume={56},
  number={8},
  pages={1--41},
  year={2024},
  doi={10.1145/3648610}
}

@article{bui2025slr,
  title={A systematic literature review on automated exploit and security test generation},
  author={Bui, Quang-Cuong and Iannone, Emanuele and Camporese, Maria and Hinrichs, Torge and Tony, Catherine and T\'{o}th, L\'{a}szl\'{o} and Palomba, Fabio and Heged\H{u}s, P\'{e}ter and Massacci, Fabio and Scandariato, Riccardo},
  journal={arXiv preprint arXiv:2502.04953},
  year={2025},
  url={https://arxiv.org/abs/2502.04953}
}

@article{liu2025webvuln,
  title={{LLM} agents for automated web vulnerability reproduction: Are we there yet?},
  author={Liu, Bin and Zhao, Yanjie and Xu, Guoai and Wang, Haoyu},
  journal={arXiv preprint arXiv:2510.14700},
  year={2025},
  url={https://arxiv.org/abs/2510.14700}
}

@article{zerodaybench2026,
  title={{ZeroDayBench}: Evaluating {LLM} agents on unseen zero-day vulnerabilities for cyberdefense},
  author={Lau, Nancy and Sloot, Louis and Raj, Jyoutir and Boscardin, Giuseppe Marco and Harris, Evan and Bowman, Dylan and Brajkovski, Mario and Chawla, Jaideep and Zhao, Dan},
  journal={arXiv preprint arXiv:2603.02297},
  year={2026},
  url={https://arxiv.org/abs/2603.02297}
}

@article{jacobs2020epss,
  title={Improving vulnerability remediation through better exploit prediction},
  author={Jacobs, Jay and Romanosky, Sasha and Adjerid, Idris and Baker, Wade},
  journal={Journal of Cybersecurity},
  volume={6},
  number={1},
  year={2020},
  doi={10.1093/cybsec/tyaa015}
}

@inproceedings{jacobs2023epssv3,
  title={Enhancing vulnerability prioritization: Data-driven exploit predictions with community-driven insights},
  author={Jacobs, Jay and Romanosky, Sasha and Suciu, Octavian and Edwards, Benjamin and Sarabi, Armin},
  booktitle={IEEE European Symposium on Security and Privacy Workshops (EuroS\&PW)},
  pages={194--206},
  year={2023},
  doi={10.1109/EuroSPW59978.2023.00027}
}

@article{gordeychik2025epss,
  title={Prediction meets patch queues: Empirical limits of {EPSS}-only prioritization using {CISA} {KEV} additions in 2025},
  author={Gordeychik, Sergey},
  journal={TechRxiv preprint},
  year={2026},
  doi={10.36227/techrxiv.176857939.95987957/v1},
}

@article{parla2024epss,
  title={Efficacy of {EPSS} in high severity {CVEs} found in {KEV}},
  author={Parla, Rianna},
  journal={arXiv preprint arXiv:2411.02618},
  year={2024},
  url={https://arxiv.org/abs/2411.02618}
}

@inproceedings{koscinski2025conflicting,
  title={Conflicting scores, confusing signals: An empirical study of vulnerability scoring systems},
  author={Koscinski, Viktoria and Nelson, Mark and Okutan, Ahmet and Falso, Robert and Mirakhorli, Mehdi},
  booktitle={Proceedings of the 2025 ACM SIGSAC Conference on Computer and Communications Security (CCS)},
  pages={1904--1918},
  year={2025},
  doi={10.1145/3719027.3765210}
}

@techreport{ssvc2026,
  title={Blueprint: {Stakeholder-Specific Vulnerability Categorization} Guidance},
  author={Smart, Justin and Jun, Sung-Yoon and others},
  institution={Sandia National Laboratories},
  year={2026},
  url={https://research-hub.nlr.gov/en/publications/blueprint-stakeholder-specific-vulnerability-categorization-guida/}
}

@techreport{nist2025lev,
  title={Likely Exploited Vulnerabilities},
  author={Mell, Peter and Spring, Jonathan},
  institution={National Institute of Standards and Technology},
  number={NIST.CSWP.41},
  year={2025},
  url={https://nvlpubs.nist.gov/nistpubs/CSWP/NIST.CSWP.41.pdf}
}

@article{alhaddad2025ssvc,
  title={Prompting the priorities: A first look at evaluating {LLMs} for vulnerability triage and prioritization},
  author={Al~Haddad, Osama and Ikram, Muhammad and Ahmed, Ejaz and Lee, Young},
  journal={arXiv preprint arXiv:2510.18508},
  year={2025},
  url={https://arxiv.org/abs/2510.18508}
}

@inproceedings{newsome2005taintcheck,
  title={Dynamic taint analysis for automatic detection, analysis, and signature generation of exploits on commodity software},
  author={Newsome, James and Song, Dawn},
  booktitle={Proceedings of the 12th Network and Distributed System Security Symposium (NDSS)},
  year={2005},
  url={https://www.ndss-symposium.org/ndss2005/dynamic-taint-analysis-automatic-detection-analysis-and-signaturegeneration-exploits-commodity/}
}

@inproceedings{liang2005covers,
  title={Fast and automated generation of attack signatures: A basis for building self-protecting servers},
  author={Liang, Zhenkai and Sekar, R.},
  booktitle={Proceedings of the 12th ACM Conference on Computer and Communications Security (CCS)},
  pages={213--222},
  year={2005},
  doi={10.1145/1102120.1102150}
}

@inproceedings{brumley2006signatures,
  title={Towards automatic generation of vulnerability-based signatures},
  author={Brumley, David and Newsome, James and Song, Dawn and Wang, Hao and Jha, Somesh},
  booktitle={Proceedings of the 2006 IEEE Symposium on Security and Privacy (S\&P)},
  pages={2--16},
  year={2006},
  doi={10.1109/sp.2006.41}
}

@article{kobayashi2023ids,
  title={Generation of {IDS} signatures through exhaustive execution path exploration in {PoC} codes for vulnerabilities},
  author={Kobayashi, Masaki and Kanemoto, Yo and Kotani, Daisuke and Okabe, Yasuo},
  journal={Journal of Information Processing},
  volume={31},
  pages={591--601},
  year={2023},
  doi={10.2197/ipsjjip.31.591}
}

@article{mitra2025falcon,
  title={{FALCON}: Autonomous cyber threat intelligence mining with {LLMs} for {IDS} rule generation},
  author={Mitra, Shaswata and Bazarov, Azim and Duclos, Martin and Mittal, Sudip and Piplai, Aritran and Rahman, Md Rayhanur and Zieglar, Edward and Rahimi, Shahram},
  journal={arXiv preprint arXiv:2508.18684},
  year={2025},
  url={https://arxiv.org/abs/2508.18684}
}

@article{shukla2025rulegenie,
  title={{RuleGenie}: {SIEM} detection rule set optimization},
  author={Shukla, Akansha and Gandhi, Parth Atulbhai and Elovici, Yuval and Shabtai, Asaf},
  journal={arXiv preprint arXiv:2505.06701},
  year={2025},
  url={https://arxiv.org/abs/2505.06701}
}

@inproceedings{vy2025evosiem,
  title={{EvoSIEM}: Detecting and generating {SIEM} rule evasion behaviors in network systems},
  author={Tran, Thi Thuy Vy and Le, Thi Bich Tuyen and Truong, Thi Hoang Hao and Thai, Hung Van and Hien, Do Hoang and Phan, The Duy},
  booktitle={IEEE RIVF International Conference on Computing and Communication Technologies},
  pages={498--503},
  year={2025},
  doi={10.1109/rivf68649.2025.11365129}
}

@inproceedings{fairbanks2025sigma,
  title={Reflective beam search for automated {TTP} extraction and Sigma rule generation from cyber threat intelligence},
  author={Fairbanks, Jeffrey and Serra, Edoardo},
  booktitle={IEEE International Conference on Big Data (BigData)},
  pages={2130--2135},
  year={2025},
  doi={10.1109/bigdata66926.2025.11401712}
}

@inproceedings{uetz2024sigmaevasion,
  title={You Cannot Escape Me: Detecting Evasions of {SIEM} Rules in Enterprise Networks},
  author={Uetz, Rafael and Herzog, Marco and Hackl{\"a}nder, Louis and Schwarz, Simon and Henze, Martin},
  booktitle={Proceedings of the 33rd USENIX Security Symposium},
  pages={5179--5196},
  year={2024},
  url={https://www.usenix.org/conference/usenixsecurity24/presentation/uetz}
}

@inproceedings{applebaum2016caldera,
  title={Intelligent, automated red team emulation},
  author={Applebaum, Andy and Miller, Doug and Strom, Blake and Korban, Chris and Wolf, Ross},
  booktitle={Proceedings of the 32nd Annual Conference on Computer Security Applications (ACSAC)},
  pages={363--373},
  year={2016},
  doi={10.1145/2991079.2991111}
}

@article{sanchezmatas2025bas,
  title={Simulating cyberattacks through a breach attack simulation ({BAS}) platform empowered by security chaos engineering ({SCE})},
  author={S\'{a}nchez-Matas, Arturo and Escribano Ruiz, Pablo and D\'{i}az-L\'{o}pez, Daniel and Perales G\'{o}mez, Angel Luis and Nespoli, Pantaleone and Mart\'{i}nez P\'{e}rez, Gregorio},
  journal={arXiv preprint arXiv:2508.03882},
  year={2025},
  url={https://arxiv.org/abs/2508.03882}
}

@article{denison2024sycophancy,
  title={Sycophancy to subterfuge: Investigating reward-tampering in large language models},
  author={Denison, Carson and MacDiarmid, Monte and Barez, Fazl and Duvenaud, David and Kravec, Shauna and Marks, Samuel and Schiefer, Nicholas and Soklaski, Ryan and Tamkin, Alex and Kaplan, Jared and Shlegeris, Buck and Bowman, Samuel R. and Perez, Ethan and Hubinger, Evan},
  journal={arXiv preprint arXiv:2406.10162},
  year={2024},
  url={https://arxiv.org/abs/2406.10162}
}

@inproceedings{yao2023react,
  title={{ReAct}: Synergizing reasoning and acting in language models},
  author={Yao, Shunyu and Zhao, Jeffrey and Yu, Dian and Du, Nan and Shafran, Izhak and Narasimhan, Karthik and Cao, Yuan},
  booktitle={Proceedings of the 11th International Conference on Learning Representations (ICLR)},
  year={2023},
  url={https://arxiv.org/abs/2210.03629}
}

\end{document}